\documentclass{article}

\usepackage{microtype}
\usepackage{graphicx}
\usepackage{subfigure}
\usepackage{booktabs} % for professional tables

\usepackage{hyperref}

\usepackage[accepted]{mlsys2025}

\mlsystitlerunning{Mapping Gemma3 onto an Edge Dataflow Architecture}

\usepackage{amsmath}
\usepackage{amsfonts}   % provides \mathbb
\usepackage{amssymb}    % loads amsfonts (also provides \mathbb)
\usepackage{multirow}
\usepackage{graphicx}
\usepackage{subcaption}
\usepackage{booktabs}
\newcommand{\github}[1]{\href{#1}{#1}} 

\begin{document}

\twocolumn[
\mlsystitle{Mapping Gemma3 onto an Edge Dataflow Architecture}

% unless you've provided the [accepted] option to the mlsys2025


\mlsyssetsymbol{equal}{*}

\begin{mlsysauthorlist}
\mlsysauthor{Shouyu Du}{equal,to}
\mlsysauthor{Miaoxiang Yu}{equal,to}
\mlsysauthor{Zhenyu Xu}{to}
\mlsysauthor{Zhiheng Ni}{to}
\mlsysauthor{Jillian Cai}{to}
\mlsysauthor{Qing Yang}{ed}
\mlsysauthor{Tao Wei}{to}
\end{mlsysauthorlist}

\mlsysaffiliation{to}{Electrical and Computer Engineering, Clemson University, USA}
\mlsysaffiliation{ed}{Electrical and Computer Engineering, University of Rhode Island, USA}

\mlsyscorrespondingauthor{Zhenyu Xu}{zxu3@clemson.edu}
\mlsyscorrespondingauthor{Tao Wei}{twei2@clemson.edu}

\mlsyskeywords{Machine Learning, MLSys}

\vskip 0.3in

\begin{abstract}
We present the first end-to-end deployment of the Gemma3 family of large language and vision models on a tiled edge dataflow architecture (AMD Ryzen AI NPU). Our work introduces a set of hardware-aware techniques. For prefill, we introduce an efficient dequantization engine, optimize tiled matrix multiplication kernels, and propose FlowQKV, a chunked, pipelined attention mechanism. For decoding, we introduce FusedDQP, which fuses dequantization and projection into a single kernel, and FlowKV, which re-structures attention to sustain high memory bandwidth utilization. Together with a compact Q4NX 4-bit quantization format, these methods yield up to \(7.5\times\) faster prefill and \(5.9\times\) faster decoding versus the iGPU, and \(23.7\times\) and \(2.7\times\) over the CPU, respectively. Power efficiency improves by as much as \(96.7\times\) and \(157.7\times\) compared to the iGPU and CPU. The proposed approach demonstrates that modern NPUs can deliver practical, low-power LLM and VLM inference at the edge, and provides a generalizable blueprint for mapping transformer-based models onto tiled dataflow accelerators.
\end{abstract}
]



\printAffiliationsAndNotice{\mlsysEqualContribution} % otherwise use the standard text.

\section{Introduction}

Local LLMs are rising for latency, offline use, and privacy: on-device inference keeps sensitive data local and avoids network delays needed for real-time apps \cite{exxact2025edgeai}. CPUs and iGPUs can be monopolized by LLM workloads, so NPUs—dedicated AI accelerators—run these tasks without degrading general-purpose performance \cite{hunhoff2025efficiency}. Major vendors (AMD, Intel, Apple, Qualcomm) now ship neural processing units (NPUs) \cite{laptopmag2024aipcs, microsoft2024npu}. AMD’s Ryzen AI NPUs use XDNA: a 2D array of AI Engine tiles forming a dataflow chip with very-long-instruction-word (VLIW) compute; their flexible data-movement fabric enables direct tile-to-tile communication, easing internal bandwidth bottlenecks and improving speed and power efficiency \cite{rico2024xdna, rosti2025unlocking, mlir_aie, hunhoff2025efficiency, 10.1145/3706628.3708870, 10.1145/3706628.3708822, 10.1145/3669940.3707239}.

However, a full end-to-end deployment of Gemma3 models (vision) on AMD Ryzen AI NPUs has not yet been demonstrated \cite{gemma3}. We present the first demonstration; and since Gemma3 is both representative and popular, the resulting insights offer a practical blueprint for subsequent model deployment and runtime designs.

LLM inference consists of two distinct phases: prefill and decoding. During prefill, the model reads the whole prompt once, stores the results as key–value (KV) cache, and--if it’s a multi-turn chat--adds these results to what was cached before, as well as image tokens (if any). This stage is compute-intensive. In contrast, decoding phase proceeds autoregressively, generating one token at a time. Each new token depends on the KV cache and previous outputs, which prevents parallelization and makes decoding memory-bound. For vision--language models (VLMs), Gemma3 incorporates a vision tower that converts an input image into 4096 tokens, which then compressed into 256 tokens, which are then processed together with text tokens during prefill.  

To accelerate prefill, we introduce a series of new techniques. First, we design an efficient dequantization engine that converts quantized projection layer weights into full precision (\texttt{bf16}). Second, we adopt a systematic, metric-driven approach to optimize tiled matrix multiplications (MMs) in \texttt{bf16}. Third, we propose FlowQKV, a novel technique designed for prefill. By exploiting dataflow, pipelining and bandwidth-aware scheduling, FlowQKV overlaps data movement with computation across multiple compute tiles (CTs), significantly improving KV cache efficiency and minimizing memory bottlenecks. One variant, FlowQKV-SWA is introduced to address Sliding Window Attention (SWA) in Gemma3. In addition, we introduce Q4NX (Quantized 4-bit NPU eXpress), a compact and NPU-friendly data format that is specifically tailored to reduce memory footprint and maximize throughput on Ryzen AI NPUs. Collectively, the optimized MM kernels, FlowQKV, and Q4NX achieve sustained high memory bandwidth utilization.  

The vision tower benefits from many of the same optimizations applied to prefill. Another FlowQKV variant,  FlowQKV-NCA (Non-Causal Attention), is introduced.

For decoding, we propose two complementary methods. The first, FusedDQP, fuses dequantization and projection into a single kernel, effectively combining them into matrix--vector multiplications (MVMs). The second, FlowKV, restructures the attention mechanism into a dataflow model, optimizing the handling of key--value pairs. As in prefill, a FlowKV-SWA variant supports efficient deployment of SWA layers. Together, FusedDQP and FlowKV maintain consistently high memory bandwidth utilization ($U_{\text{mem}}^{rd}$) during decoding, thereby improving latency and throughput.  

We evaluate 4-bit quantized Gemma3 models (1B and 4B) on an AMD Ryzen AI 7 350 NPU. Using the same platform’s iGPU and CPU as baselines, the NPU achieves, for prefill, up to \(7.5\times\) over the iGPU and \(23.7\times\) over the CPU. For decoding—especially at longer sequences—the NPU matches or exceeds both baselines, reaching up to \(5.9\times\) over the iGPU and \(2.7\times\) over the CPU. For the vision tower (processing and understanding; 4B only), the NPU is \(2.9\times\) faster than the iGPU and \(14.8\times\) faster than the CPU. Importantly, the NPU demonstrates markedly superior energy efficiency—achieving up to \(96.7\times\) and \(157.7\times\) improvements over the iGPU and CPU, respectively, as measured in Tokens Per Second per Watt (TPS/W).

Main contributions of this work are as follows:
\begin{itemize}
  \item We present the \textit{first end-to-end deployment} of Gemma3 models on an edge dataflow architecture.
  \item We propose a set of \textit{novel techniques} to enable efficient prefill, decoding, and vision-tower execution.
  \item We share the \textit{lessons learned} from this deployment, highlighting opportunities and challenges.
\end{itemize}
\section{Background}
\label{sec:background}
This section briefs the computer architecture of the Ryzen AI NPU and model architecture of Gemma3.

\subsection{AMD Ryzen AI NPU Architecture}
\label{sec:NPU_arch}

\subsubsection{Platform Overview}

\begin{figure}[h!]
    \centering
    \includegraphics[width=0.30\textwidth]{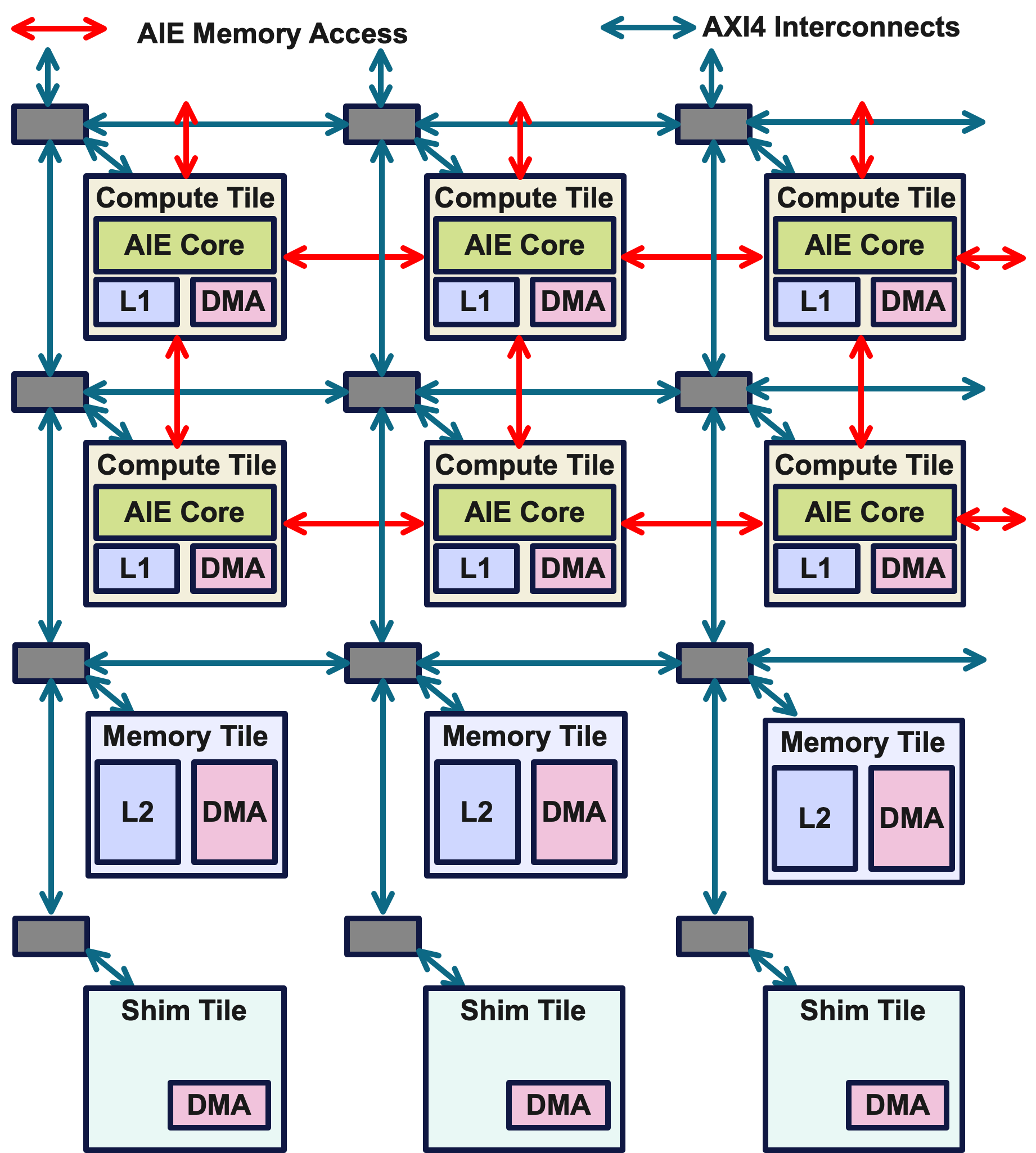} % Replace 'example-image' with your image file name
    \caption{NPU architecture overview}
    \label{fig:npu_arch}
    % \Description{NPU architecture overview.}
\end{figure}

The AMD Ryzen AI NPU architecture is built around interconnected 2D processing units \cite{rico2024xdna}. Exemplified by the NPU2 (XDNA2), these units are referred to as Compute Tiles (CTs). It features a 32-CT grid (8 columns and each contains 4 CTs) in a checkerboard layout, shown in Fig. \ref{fig:npu_arch}. These instructions are dispatched across two parallel processors: the scalar processor, which handles complex pointer arithmetic, and the vector processor, which is optimized for SIMD (Single Instruction Multiple Data) operations. Each CT runs at a clock speed up to 1.8 GHz.

\subsubsection{Compute Tiles (CTs)}
\label{sec:CT}
\begin{figure}[!ht]
    \centering
    \includegraphics[width=0.35\textwidth]{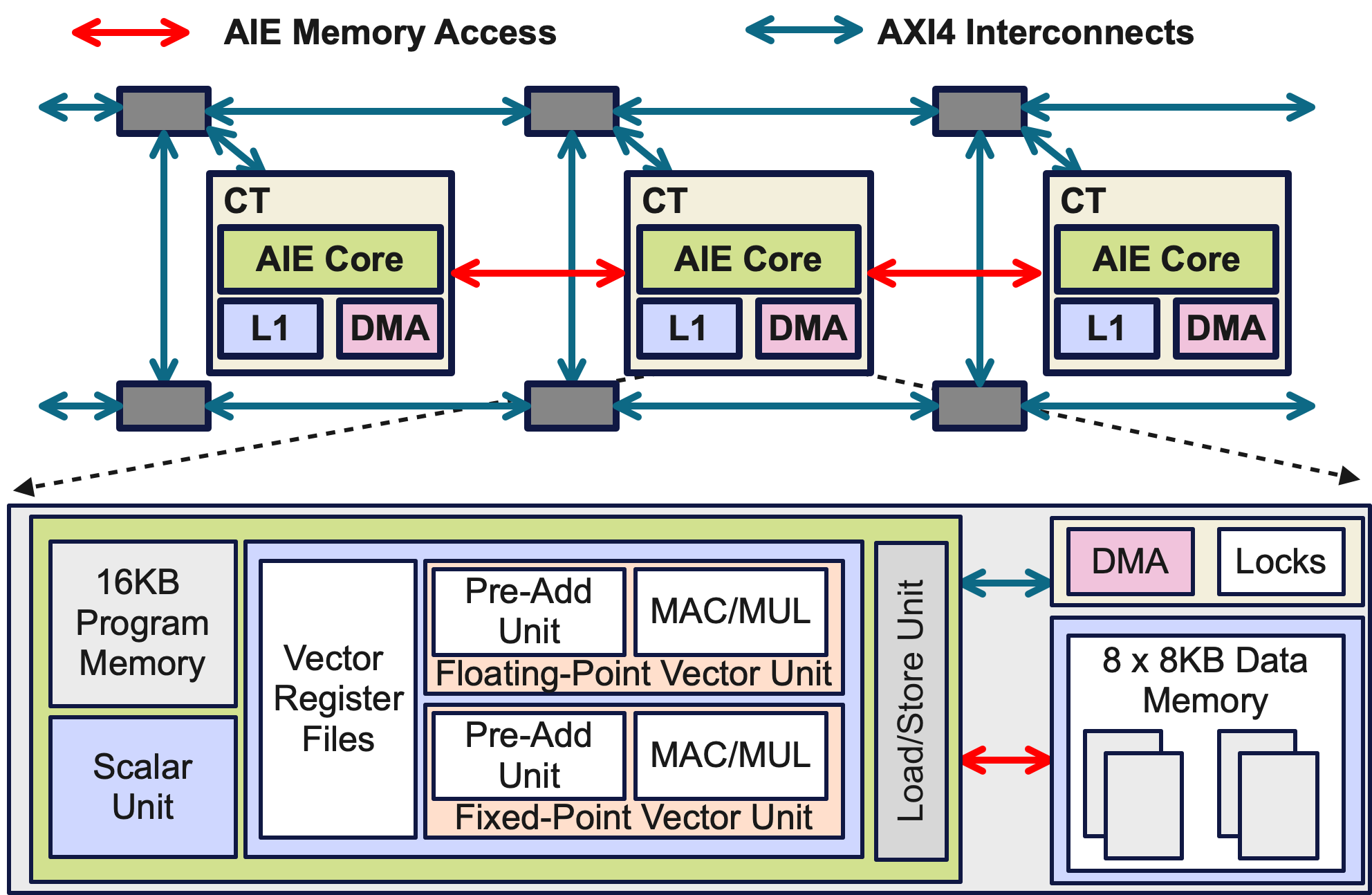} % Replace 'example-image' with your image file name
    \caption{CT overview}
    \label{fig:ct_arch}
    % \Description{CT overview}
\end{figure}

The vector processor supports both floating-point and fixed-point operations. Integrated pre-add units handle basic vector functions such as minimum/maximum calculations and direct vector comparisons. CT local data memory (DM), also referred to as L1 memory, is divided into eight banks (a total of 64 KB), each accessible via a DMA (Direct Memory Access) interface, shown in Fig. \ref{fig:ct_arch}.

% Synchronization locks are implemented to manage coordination across CT-to-CT operations, CT-DMA interactions, and connections between CTs and external memory-mapped devices.

CTs communicate through two primary interfaces. The first is AIE (AI Engine) memory access, enabling direct load/store operations between neighboring CTs. The second is the AXI4 (Advanced eXtensible Interface 4) interconnect, which allows DMA transfers to non-adjacent CTs via AXI4 streams. This interface also supports input broadcasting by duplicating data for multiple CTs.

\subsubsection{Memory Tiles (MTs) and Shim Tiles (STs)}

The platform includes 8 Memory Tiles (MTs), positioned at the bottom of the CT array, shown in Fig. \ref{fig:npu_arch}. Each tile contains 512 KB of high-density, high-bandwidth memory, also referred to as L2 memory. Shim Tiles (STs) provide communication with access to main memory (off-chip DDR memory). MT interfaces with ST to access main memory via AXI4 streams. Lightweight DMA engines within the MT and ST ensure efficient data transfer between the main memory and the CT array.

\subsubsection{Programming}

% AMD Ryzen AI NPU integrates independent and scalable datapaths and robust memory management to enable high-performance parallel computations. 
Programming the platform relies on two main components. Kernels are directly mapped to CTs to handle core computations, while the Multi-Level Intermediate Representation (MLIR) manages connections and creates data buffers between tiles, facilitating efficient communication and coordination \cite{mlir_aie}. These features collectively ensure scalability and the efficient execution of large-scale computations.
\begin{figure}[!ht]
    \centering
    \includegraphics[width=0.3\textwidth]{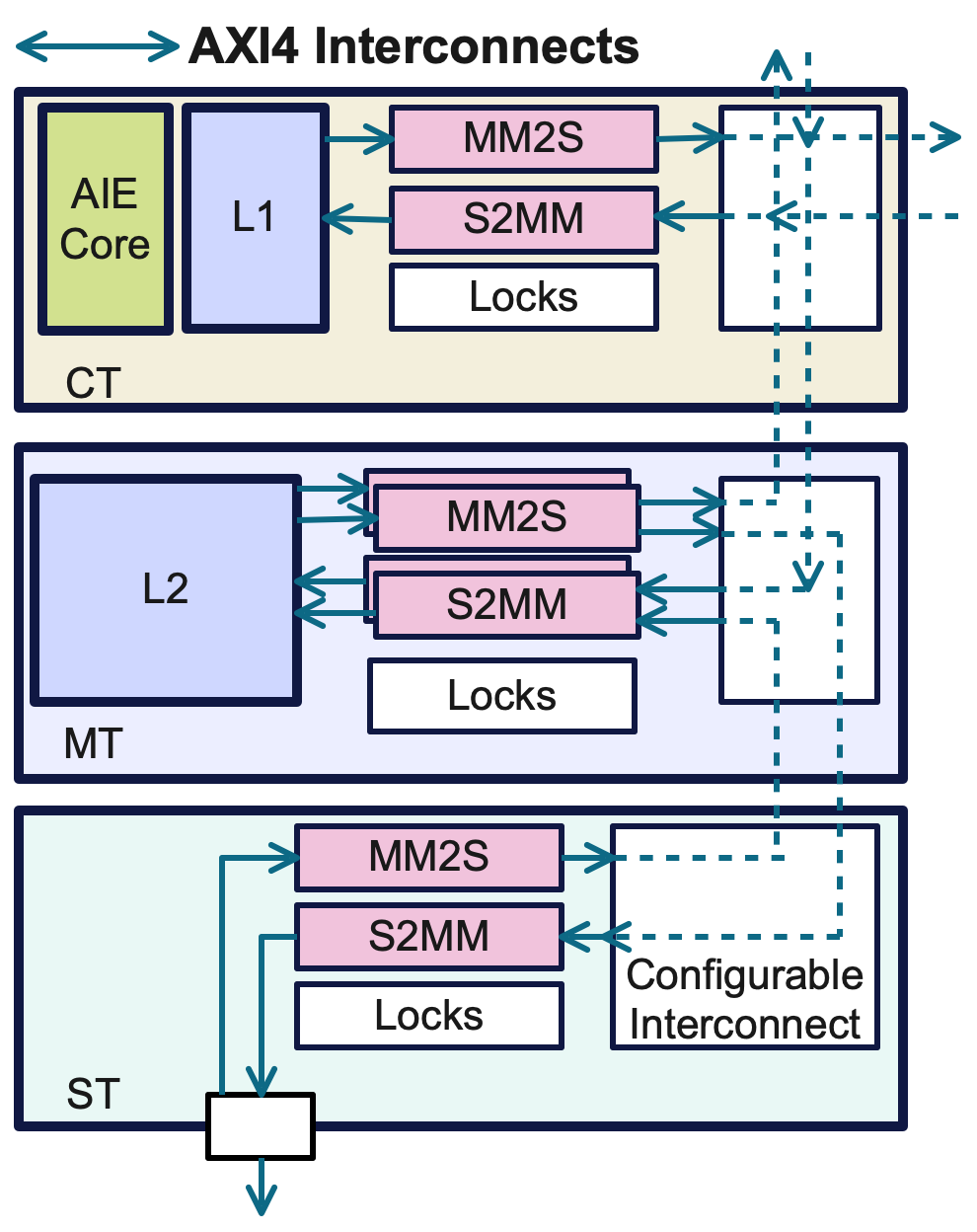} % Replace 'example-image' with your image file name
    \caption{Data movement architecture on AMD NPU}
    \label{fig:data_movement}
    % \Description{Data movement architecture on AMD NPU}
\end{figure}

\subsubsection{Data Movement on NPU}

% To optimize application performance, the data movement architecture ensures efficient transfer of inputs and outputs between the core, L1, L2, and main memory. These transfers must occur at sufficient speeds to keep the core fully utilized.

The primary data transfers occur between main memory and L2, as well as between L2 and L1. Additionally, the architecture supports flexible interconnects, enabling direct transfers such as main memory to L1 and L1 to L1. A data transfer involves a source DMA channel (MM2S, or Memory-Mapped to Stream), a destination DMA channel (Steam to Memory-Mapped), and a configured stream interconnect connecting the two (see Fig. \ref{fig:data_movement}). The MM2S channel reads data from memory and places it on the stream, while the S2MM channel retrieves the data from the stream and writes it to memory. Each CT, MT, and ST is equipped with a stream interconnect switch and multiple MM2S and S2MM DMA channels (AXI4 interconnects). In the ST, DMA channels read and write to the main memory. Additionally, double buffering of input and output buffers in L1 memory allows data transfers to overlap with computation, effectively hiding data transfer latency.

\subsection{Model Architecture of Gemma3}

\subsubsection{Model Architecture}

\begin{figure}[!ht]
    \centering
    \includegraphics[width=0.5\textwidth]{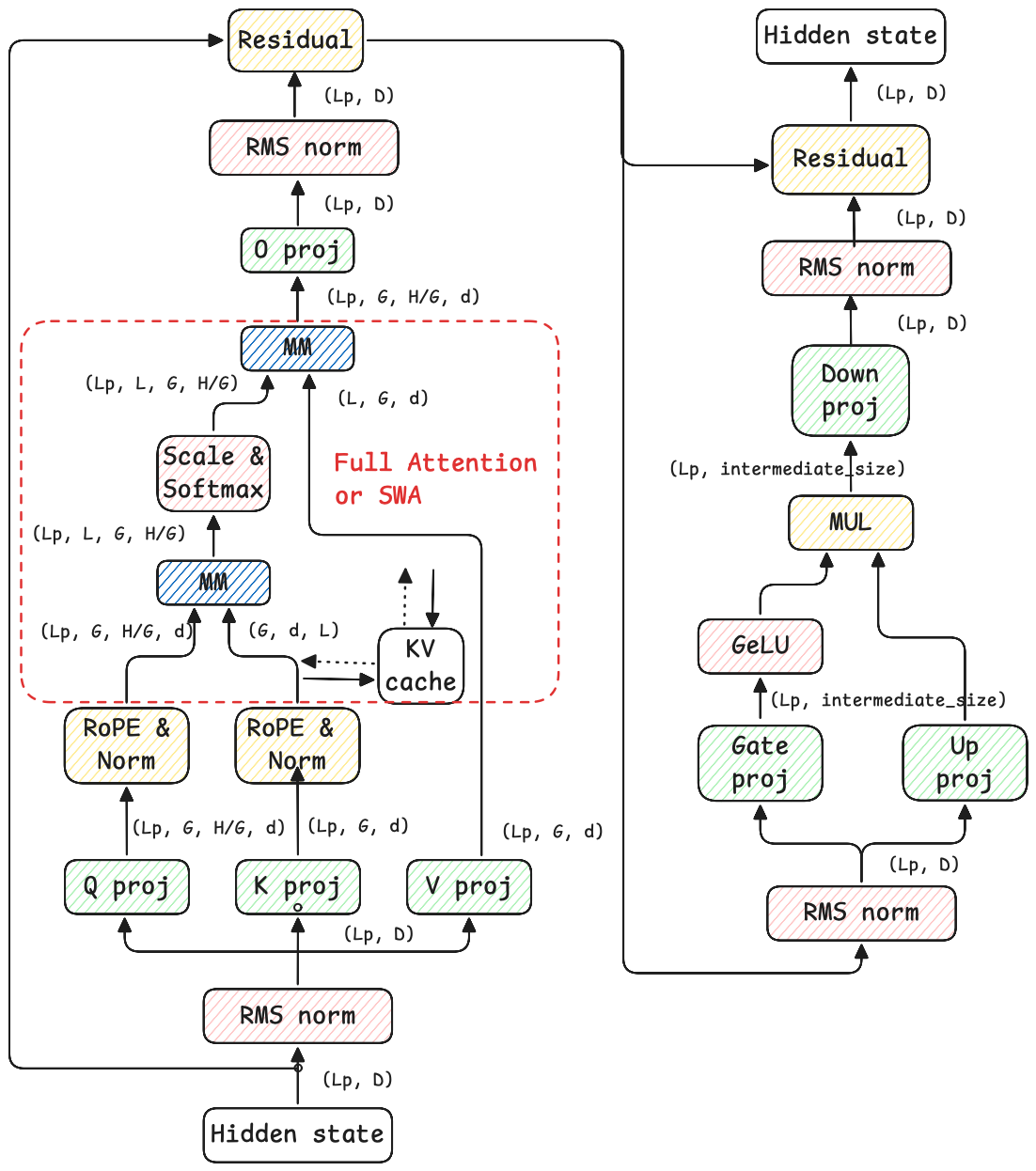} % Replace 'example-image' with your image file name
    \caption{Gemma3 4B model architecture (text portion; one transformer layer shown).
    Key model parameters: $D = 2560$ is the model dimension, $H = 8$ is the number of attention heads, and $G = 4$ is the number of KV groups. 
    The head dimension $d = 256$. 
    This $d$ is the dimensionality of each query ($q$), key ($k$), and value ($v$) vector per head.
    In prefill, $L_p$ represents the sequence length (token length of the current prompt).
    In decoding, $L_p = 1$.
    $L$ represents the total sequence length, which increases progressively during autoregressive LLM inference (In local layer, the window length, $L_w$, for SWA is 1024).
    The attention module (both full and SWA) is highlighted inside the dotted \textcolor{red}{red} box.
    }
    \label{fig:Gemma3_4B}
\end{figure}

The model (text portion) consists of 34 transformer layers, with a pattern of 5 local layers (SWA with a window size of 1024) for every global layer (full attention), starting with a local layer as the first layer of the model \cite{gemma3}; one representative layer is shown in Fig.~\ref{fig:Gemma3_4B}, with dimensions and notation defined in the caption. 

The vision tower is a 400M-parameter SigLIP Vision Transformer (ViT) \cite{zhai2023sigmoid}. During preprocessing, each image is tokenized into 4,096 tokens, which are processed by 24 transformer layers (full, non-causal attention, no grouped query attention, or GQA). Functionally, this stage can be viewed as prefill, producing visual context (256 tokens per image) that seeds the language stack.

At a high level, the Gemma3 architecture (text and vision) comprises three operation classes: (i) projection operations / matrix multiplications (MM); (ii) attention computation—full or sliding-window (SWA), causal or non-causal—highlighted by the dotted red box in Fig.~\ref{fig:Gemma3_4B}; and (iii) nonlinear functions such as RoPE, RMS normalization, GeLU, and QK-Norm.

\subsubsection{Projection Operation during Prefill--MM}
Projection weights $\mathbf{W}$ (e.g., $\mathbf{W}_q$, $\mathbf{W}_k$, $\mathbf{W}_v$, and $\mathbf{W}_o$) are applied to the input matrix. On edge devices, these weights are typically stored in low-precision formats--most commonly 4-bit integers (\texttt{int4})--to alleviate off-chip memory capacity and memory bandwidth constraints. During prefill, the weights are first dequantized to full precision (\texttt{bf16}), and then transferred to the dedicated compute unit (e.g., NPU) to perform the MM operation. Note that in the vision tower, MM is handled similarly, except no quantization is used.

\subsubsection{Attention Computation during Prefill}

In the prefill phase, the model makes a single forward pass over the full prompt, stores the results as a KV cache, and--when it’s a multi-turn conversation--appends these results to the prior cache, as well as image tokens (if any). This passes through all layers (some are SWA in Gemma3, which lowers compute intensity). This prepares the model to generate the first output token, which then initiates the decoding phase.

To simplify the explanation, we consider a single-head attention mechanism and temporarily ignore GQA. Let \( L \) denote the total context length, including both the previous tokens and the current prompt, and let \( L_p \) denote the number of tokens in the current prompt (i.e., the new input being prefed). We define the query matrix \( \mathbf{Q} \in \mathbb{R}^{L_p \times d} \), the key matrix \( \mathbf{K} \in \mathbb{R}^{L \times d} \), and the value matrix \( \mathbf{V} \in \mathbb{R}^{L \times d} \). The attention output matrix \( \mathbf{O} \in \mathbb{R}^{L_p \times d} \) is computed using the scaled dot-product attention formula:

\begin{equation}
\mathbf{O} = \text{softmax}\left( \frac{\mathbf{Q} \mathbf{K}^\top}{\sqrt{d}} \right) \mathbf{V}
\label{eq:matrix_attention}
\end{equation}

Here, the product \( \mathbf{Q} \mathbf{K}^\top \in \mathbb{R}^{L_p \times L} \) contains the raw attention scores between each new query token and all tokens in the context. The softmax function is applied row-wise over the sequence length \( L \), producing a normalized attention weight matrix. The final output \( \mathbf{O} \) is obtained by applying these attention weights to the value matrix \( \mathbf{V} \).

In Gemma3 (text portion), GQA is used, where each KV group is shared among \( H/G \) query heads. H represents the number of attention heads and G represents the number of KV groups. As the conversation progresses and the total sequence length increases, the number of stored keys and values grows linearly with \( L \), resulting in increased memory access and compute demand. This makes attention computation the dominant performance bottleneck, particularly at long sequence lengths. Accordingly, Gemma-3 introduces local layers (via SWA) that attend only to the most recent \(L_w\) tokens. In addition, note that in the vision tower, attention is handled similarly, except it uses full non-causal attention.

\subsubsection{Projection Operation during Decoding--MVM}
During LLM decoding, the weights are first transferred to the compute unit, then dequantized back to \texttt{bf16} precision before projection is performed. This introduces two distinct operations: dequantization and MVM. The separation of these steps incurs additional memory access and latency overhead, which becomes a performance bottleneck during projection. Addressing this inefficiency is critical for optimizing LLM decoding.

\subsubsection{Attention Computation during Decoding}
Shown in Eq. \ref{eq:causal_attention}, the attention output at decoding step \( t \), denoted as \( \mathbf{o}_t \in \mathbb{R}^d \), is computed as a weighted sum over the value vectors \( \mathbf{v}_j \in \mathbb{R}^d \) from all previous time steps \( j = 1 \) to \( t \). Specifically, the attention weight for each position \( j \) is determined by the dot product between the current query vector \( \mathbf{q}_t \in \mathbb{R}^d \) and each key vector \( \mathbf{k}_j \in \mathbb{R}^d \), followed by softmax normalization over all previous positions. For simplicity, the scaling factor \( \sqrt{d} \) is omitted here. This produces a probability distribution over prior tokens, reflecting their relative importance to the current token. The full attention computation is given by:
\begin{equation}
\mathbf{o}_t = \sum_{j=1}^{t} \frac{\exp(\mathbf{q}_t^\top \mathbf{k}_j)}{\sum_{l=1}^{t} \exp(\mathbf{q}_t^\top \mathbf{k}_l)} \mathbf{v}_j \quad \in \mathbb{R}^d.
\label{eq:causal_attention}
\end{equation}
Here, \( d \) denotes the dimension of each query, key, and value vector. Gemma3 uses GQA, where $H/G$ queries in a KV group share 1 KV group. As decoding proceeds, the number of stored keys and values increases linearly with \( t \), leading to growing memory access and compute requirements. This makes attention computation the primary performance bottleneck at longer sequence lengths. Note that in local layers, SWA is applied and the summation range in Eq.~\ref{eq:causal_attention} is restricted to \(t - L_w + 1 \ldots t\) when \(t > L_w\) (otherwise \(1 \ldots t\)).

\subsubsection{Nonlinear functions}
The computational cost of nonlinear functions in both prefill and decoding phases is comparatively low. Thus, our design will focus on projection and attention operations.

\section{Working Principles}

\subsection{Prefill}

This section presents optimization techniques specifically designed for the NPU architecture. 

% The first is an efficient dequantization engine. The second is a metric-driven method for optimizing tiling-based MM, tailored to the memory hierarchy and data movement constraints of the NPU. The third, called FlowQKV, is a pipelined, hardware-aware attention mechanism that improves the efficiency of attention computation by maximizing throughput and minimizing memory access latency across CTs.

\subsubsection{Dequantization Engine}

Q4\_1 quantization—using \texttt{int4} weights and \texttt{bf16} activations—is widely adopted for deploying LLMs on edge devices due to its favorable trade-off between model size and accuracy. Here, we use \texttt{bf16} for all activations and non-projection weights in the proposed design. The dequantization logic is modular and format-agnostic, supporting schemes like Q4\_1 or any block-wise quantization. In these schemes, weights are quantized in groups (e.g., size 32, 64, 128), each with a shared scale factor and minimal value offset, following:
\begin{equation}
\hat{w}_i = d_g \cdot w_{qi} + m_g
\label{eq:dequant}
\end{equation}
where $w_{qi} \in \{0, \dots, 15\}$ is the \texttt{int4} weight, $d_g \in \mathbb{R}$ is the scale, and $m_g \in \mathbb{R}$ is the minimal value offset of group $g$.

We adopt group size $g = 32$, the minimum of commonly used group sizes, for best compatibility. As on AMD Ryzen AI NPU2, only \textbf{bf16} precision multiplication is natively supported, all minimal value offsets are pre-converted to \textbf{bf16} offline. To reduce on-chip memory usage, projection weights are tiled into blocks with a size of $32 \times 256$. Each block stores \(32\times256\) \texttt{int4} weights, \(256\) \texttt{bf16} scales, and \(256\) \texttt{bf16} minimal value offset, for a total size of \(5{,}120\) bytes (\(5.0\) KB). We refer to this compact format as Q4NX (Quantized 4-bit NPU eXpress). Note that Q4NX can be extended to support emerging MXFP4, making it future-proof.

Each CT processes a Q4NX block by dequantizing it using Eq.~\ref{eq:dequant} and writing the full-precision result to DDR. Input (quantized) and output (dequantized) data are double-buffered to overlap computation and transfer.

% Let $\text{CT}_{\text{count}}$ be the number of CTs, $\text{block\_size}$ the number of weights per block, and $t_d$ the time to dequantize one block. The effective memory bandwidth utilizations are:

% \begin{equation}
% U_{\text{mem}}^{rd} = \frac{\text{CT}_{\text{count}} \times 5~\text{bits} \times \text{block\_size}}{t_d}
% \label{eq:dequant_read_bw}
% \end{equation}

% \begin{equation}
% U_{\text{mem}}^{wr} = \frac{\text{CT}_{\text{count}} \times 2~\text{bytes} \times \text{block\_size}}{t_d}
% \label{eq:dequant_write_bw}
% \end{equation}

\subsubsection{Optimizing Projection Operations}

In the prefill, all projection operations are implemented as MMs. This section begins by reviewing the fundamentals of tiling-based MM, followed by its mapping and implementation on AMD Ryzen AI NPUs. We then introduce performance metrics that guide the optimization process to improve compute efficiency and memory utilization.

\begin{figure}[!ht]
    \centering
    \includegraphics[width=0.5\textwidth]{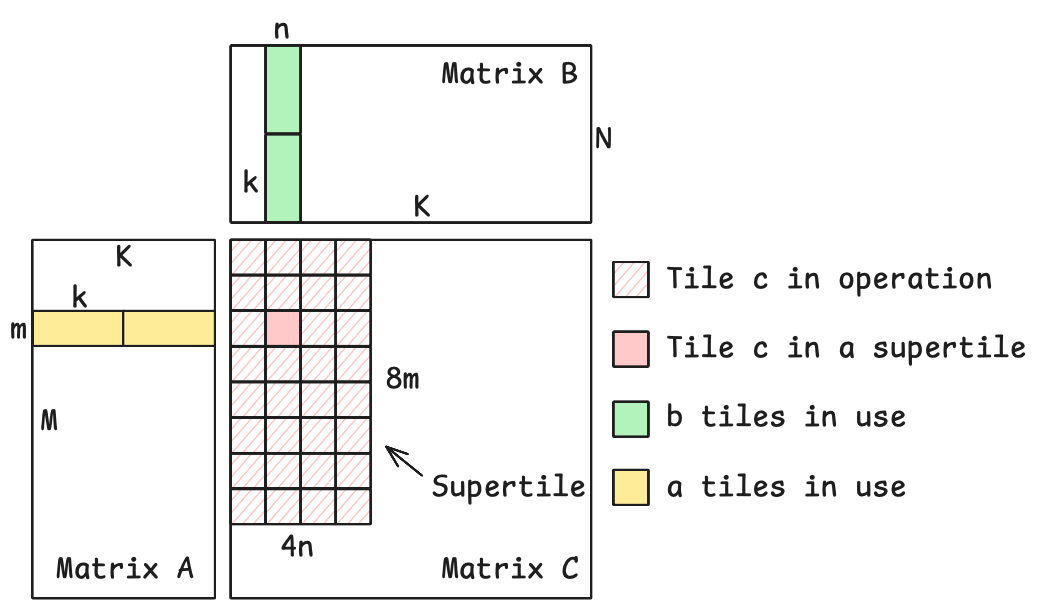} % Replace 'example-image' with your image file name
    \caption{Diagram of tiling-based MM. The concept of a supertile is specific to the AMD Ryzen AI NPU2 implementation and reflects the aggregated output block computed in $K/k$ load-compute cycles (2 in this example).}
    \label{fig:MM}
\end{figure}

Tiling-based MM is a well-established optimization technique for improving data locality and compute efficiency \cite{goto2008anatomy}. Consider the standard MM \( \mathbf{C} = \mathbf{A} \times \mathbf{B} \), where \( \mathbf{A} \in \mathbb{R}^{M \times K} \), \( \mathbf{B} \in \mathbb{R}^{K \times N} \), and \( \mathbf{C} \in \mathbb{R}^{M \times N} \) (Fig.~\ref{fig:MM}). In the tiling-based approach, matrices \( \mathbf{A} \) and \( \mathbf{B} \) are partitioned into smaller submatrices (tiles) of dimensions \( m \times k \) and \( k \times n \), respectively. Each output tile \( \mathbf{c} \in \mathbb{R}^{m \times n} \) is computed from the corresponding tiles \( \mathbf{a} \) and \( \mathbf{b} \). Once all output tiles are computed and accumulated, the full matrix \( \mathbf{C} \) is assembled. The total number of floating-point operations (multiply and add) required is \( 2 \cdot M \cdot K \cdot N \). Assuming a single compute unit and ignoring reuse across tiles, the total data movement cost is approximately:
\begin{equation}
\frac{N}{n} \cdot \text{size}(\mathbf{A}) + \frac{M}{m} \cdot \text{size}(\mathbf{B})
\label{eq:data_movement}
\end{equation}
This reflects the cost of loading the necessary tiles of \( \mathbf{A} \) and \( \mathbf{B} \) into local memory to compute all tiles of \( \mathbf{C} \).

\begin{figure}[!ht]
    \centering
    \includegraphics[width=0.4\textwidth]{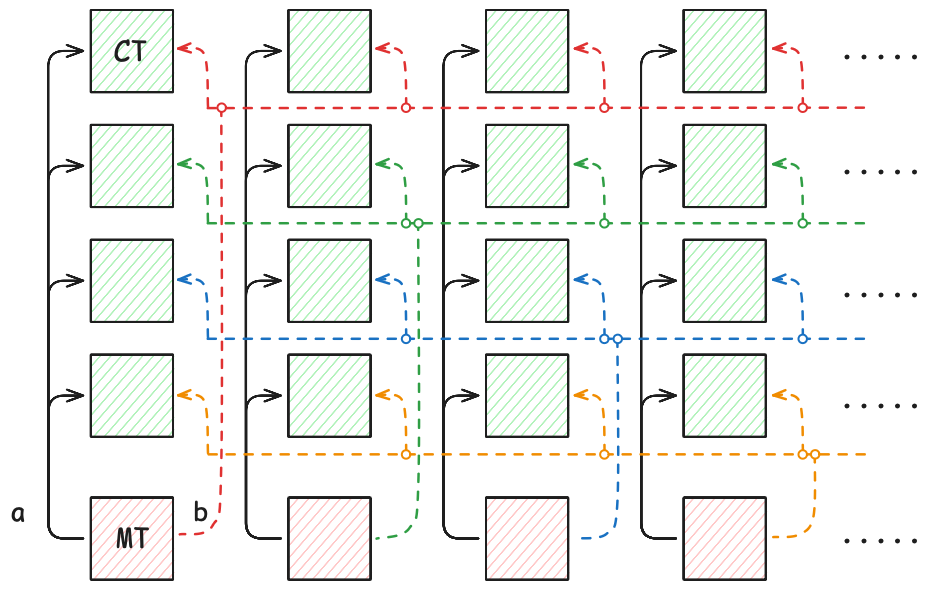} % Replace 'example-image' with your image file name
    \caption{Dataflow diagram of tiling-based MM on NPUs. Solid black arrows indicate the broadcast of tile \( \mathbf{a} \) to all CTs within a column, while dotted black arrows represent the broadcast of tile \( \mathbf{b} \) to all CTs within a row.} 
    \label{fig:MM_NPU}
\end{figure}

Efficient tiling-based MM on NPUs has been previously studied~\cite{mlir_aie, rosti2025unlocking}. In this approach, input matrices \( \mathbf{A} \) and \( \mathbf{B} \) are distributed to compute tiles (CTs) via DMA channels, as illustrated in Fig.~\ref{fig:MM_NPU}. Specifically, tiles of \( \mathbf{A} \) (denoted \( \mathbf{a} \)) are broadcast to all CTs within a column, while tiles of \( \mathbf{B} \) (denoted \( \mathbf{b} \)) are broadcast to all CTs within a row.

The compute array is \(8\times4\) CTs; each column receives a unique \( \mathbf{a} \) tile and each row a unique \( \mathbf{b} \) tile. Tiles load concurrently via DMA with double buffering to overlap transfer and compute, and the resulting \( \mathbf{c} \) tiles are written back to DDR (write-back paths omitted).

This dataflow computes an effective block of the output matrix \( \mathbf{C} \) with dimensions \( 8m \times 4n \) in $K/k$ load-and-compute cycles, shown in Fig.~\ref{fig:MM}. We refer to this block as a \textit{supertile}.

When the read memory bandwidth \( U^{rd}_{\text{mem}} \) is higher that the allocated memory bandwidth (memory-bound), we introduce a \textit{megatile}, which is formed on the MT (L2). For example, if a supertile is sized \( 8m \times 4n \), then a megatile spans \( 8e_y \cdot m \times 4e_x \cdot n \), where \( e_x \) and \( e_y \) are expansion factors along the \( x \)- and \( y \)-axes, respectively. As a complete megatile (output \( C \) block) surpasses the aggregate L1 memory capacity of the CTs, it must be staged and transferred out through the MTs (L2). The total data movement cost associated with megatile processing is given by:

\begin{equation}
\frac{N}{4e_x n} \cdot \text{size}(\mathbf{A}) + \frac{M}{8e_y m} \cdot \text{size}(\mathbf{B})
\label{eq:mega_data_movement_npu}
\end{equation}

We evaluated megatile shapes \(128\times512\times512\), \(256\times256\times512\), and \(512\times512\times512\) on the Kraken Point NPU, achieving \(5.9\), \(12.0\), and \(13.7\) TOPS, respectively (measured on a \(2048\times2048\times2048\) \texttt{bf16} GEMM). Larger tiles improve memory-bandwidth efficiency and speed, but real workloads have varying sequence lengths ($M$): shorter inputs require more zero-padding, reducing efficiency. Smaller tiles reduce padding but lower throughput. So, we decouple \(e_y m\) and \(8e_y m\), leading to a minimal size of \((e_y m)\times(4 e_x n)\) (when only 1 CT column is in use) for minimum zero-padding when sequence length ($M$) is small.

\subsubsection{FlowQKV}
\label{sec:FlowQKV}

FlowQKV is a hardware-aware attention mechanism optimized for the prefill phase. It addresses the bandwidth bottleneck caused by full-sequence attention by restructuring how the model accesses and processes KV cache data. The method is explicitly designed to maximize DRAM bandwidth efficiency and enable pipelined execution across CTs.

During prefill, attention must use the current query matrix \( \mathbf{Q} \) and all past KV cache entries accumulated across multiple conversation turns. This makes attention computation memory-intensive, especially for long sequences, as every query token must attend to all previous tokens. To address this, we apply a chunked processing strategy.

The input sequence is divided into fixed-size chunks of length \( L_c \), and attention is computed incrementally across these chunks using a numerically stable accumulation scheme~\cite{milakov2018online, dao2022flashattention}. Crucially, this approach preserves \textit{causal semantics}: each query chunk attends to all key-value tokens up to and including the current chunk.

For chunked attention, let
$\mathbf{Q}_c\!\in\!\mathbb{R}^{L_c\times d}$ be the query matrix for the current chunk $c$;
$\mathbf{K}_i,\mathbf{V}_i\!\in\!\mathbb{R}^{L_c\times d}$ be keys/values for KV-chunk $i\le c$ (context up to and including $c$);
$\mathbf{Y}_{\text{left}}\!\in\!\mathbb{R}^{L_c\times d}$ be the accumulated softmax-weighted numerators from prior chunks; and
$\mathbf{l}_{\text{left}}\!\in\!\mathbb{R}^{L_c\times 1}$ be the corresponding accumulated denominators.
These accumulators enable numerically correct, memory-efficient incremental attention across chunks, well-suited to tiled NPUs.

For clarity, we consider full causal attention for a single KV head (i.e., without GQA). At each chunk step, attention is computed as follows:

\begin{enumerate}
  \item Attention Score Computation:
  \begin{equation}
    \mathbf{S}_c = {\mathbf{Q}_c \mathbf{K}_c^\top}/{\sqrt{d}}
    \label{eq:gqa_score}
  \end{equation}
  where \( \mathbf{S}_c \in \mathbb{R}^{L_c \times L_c} \) contains the raw attention scores between current queries and keys in this chunk.

  \item Numerical Stabilization:
  \begin{equation}
    \mathbf{m}_c = \max(\mathbf{S}_c, \mathbf{m}_c^{\text{left}})
    \label{eq:gqa_max}
  \end{equation}
  where \( \mathbf{m}_c, \mathbf{m}_c^{\text{left}} \in \mathbb{R}^{L_c \times 1} \) are row-wise max values across current and previous chunks.

  \item Exponentiation with Shift:
  \begin{equation}
    \mathbf{F}_c = \exp(\mathbf{S}_c - \mathbf{m}_c)
    \label{eq:gqa_exp}
  \end{equation}
  where \( \mathbf{F}_c \in \mathbb{R}^{L_c \times L_c} \) contains the softmax numerators.

  \item Correction Factor for Previous Chunks:
  \begin{equation}
    \mathbf{C}_c = \exp(\mathbf{m}_c^{\text{left}} - \mathbf{m}_c)
    \label{eq:gqa_correction}
  \end{equation}
  where \( \mathbf{C}_c \in \mathbb{R}^{L_c \times 1} \) ensures proper scaling when combining current and prior accumulations.

  \item Accumulate Softmax Denominator:
  \begin{equation}
    \mathbf{l} = \mathbf{C}_c \bigodot \mathbf{l}_{\text{left}} + \sum \mathbf{F}_c \quad \text{(row-wise sum)}
    \label{eq:gqa_denominator}
  \end{equation}
  where \( \mathbf{l}, \mathbf{l}_{\text{left}} \in \mathbb{R}^{L_c \times 1} \) represent the running softmax denominator across all chunks seen so far.

  \item Accumulate Softmax-Weighted Values:
  \begin{equation}
    \mathbf{Y} = \mathbf{C}_c \bigodot \mathbf{Y}_{\text{left}} + \mathbf{F}_c \mathbf{V}_c
    \label{eq:gqa_numerator}
  \end{equation}
  where \( \mathbf{Y}, \mathbf{Y}_{\text{left}} \in \mathbb{R}^{L_c \times d} \) accumulate the weighted value vectors across chunks.

  \item Final Attention Output (after all chunks):
  \begin{equation}
    \mathbf{O} = \mathbf{Y}/\mathbf{l}
    \label{eq:gqa_output}
  \end{equation}
  where \( \mathbf{O} \in \mathbb{R}^{L_c \times d} \) is the final attention output for the current query chunk.
\end{enumerate}

This chunked attention approach, illustrated in Fig.~\ref{fig:chunked_attention}, offers several key advantages critical for efficient execution on hardware-constrained platforms like AMD Ryzen AI NPUs:
\begin{figure}[!ht]
    \includegraphics[width=0.48\textwidth]{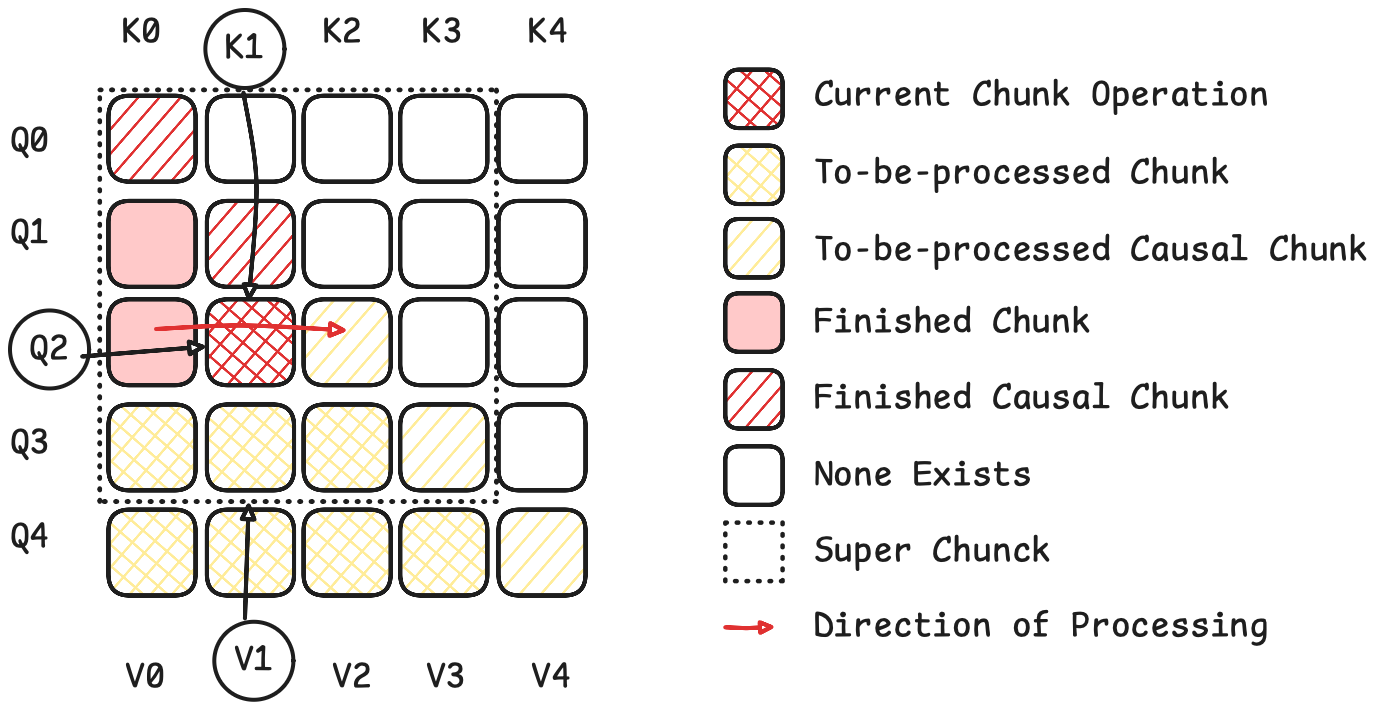} % Replace 'example-image' with your image file name
\caption{
Illustration of the chunked attention workflow. Processing proceeds strictly from left to right, as indicated by the red arrow, to preserve causal attention. Chunks along the diagonal contain only lower triangular regions, reflecting the causal constraint that tokens cannot attend to future positions. Each chunk operation requires the corresponding \( \mathbf{Q} \), \( \mathbf{K} \), and \( \mathbf{V} \) blocks (e.g., \( \mathbf{Q_2} \), \( \mathbf{K_1} \), and \( \mathbf{V_1} \) here), as well as the intermediate values \( \mathbf{m}_c^{\text{left}} \), \( \mathbf{l}_{\text{left}} \), and \( \mathbf{Y}_{\text{left}} \) passed from the preceding (left) chunk.
}
\label{fig:chunked_attention}
\end{figure}

% \begin{itemize}
%   \item \textbf{Causal attention is preserved:} Each query token in the current chunk attends to all key-value vectors from both the current and all previous chunks. Shown in Fig.~\ref{fig:chunked_attention}, mask is required for causal chunks (diagonal or end chunk for each process).

%   \item \textbf{Bounded memory usage:} Instead of storing the full softmax numerator and denominator matrices across the entire sequence, only the running partial accumulations—\( \mathbf{Y}_{\text{left}} \) and \( \mathbf{l}_{\text{left}} \)—are retained. 

%   \item \textbf{Hardware-efficient tiling:} By processing one chunk at a time, the compute and memory requirements of each attention block are kept small enough to fit within the a CT, enabling pipelined execution with double-buffered DMA transfers, allowing compute and memory access to overlap, thereby maximizing throughput.
% \end{itemize}
In chunked attention, \emph{causality is preserved} by allowing each query in the current chunk to attend to all K/V from the current and all prior chunks, with a causal mask (diagonal or end-chunk per process) as in Fig.~\ref{fig:chunked_attention}; \emph{memory remains bounded} by keeping only the running accumulations \( \mathbf{Y}_{\text{left}} \) and \( \mathbf{l}_{\text{left}} \) instead of full softmax numerators/denominators; and \emph{hardware efficiency} is achieved via chunkwise tiling that fits each attention block within a compute tile (CT) and uses double-buffered DMA to overlap transfer and compute, maximizing throughput.

\begin{figure}[!ht]
    \includegraphics[width=0.48\textwidth]{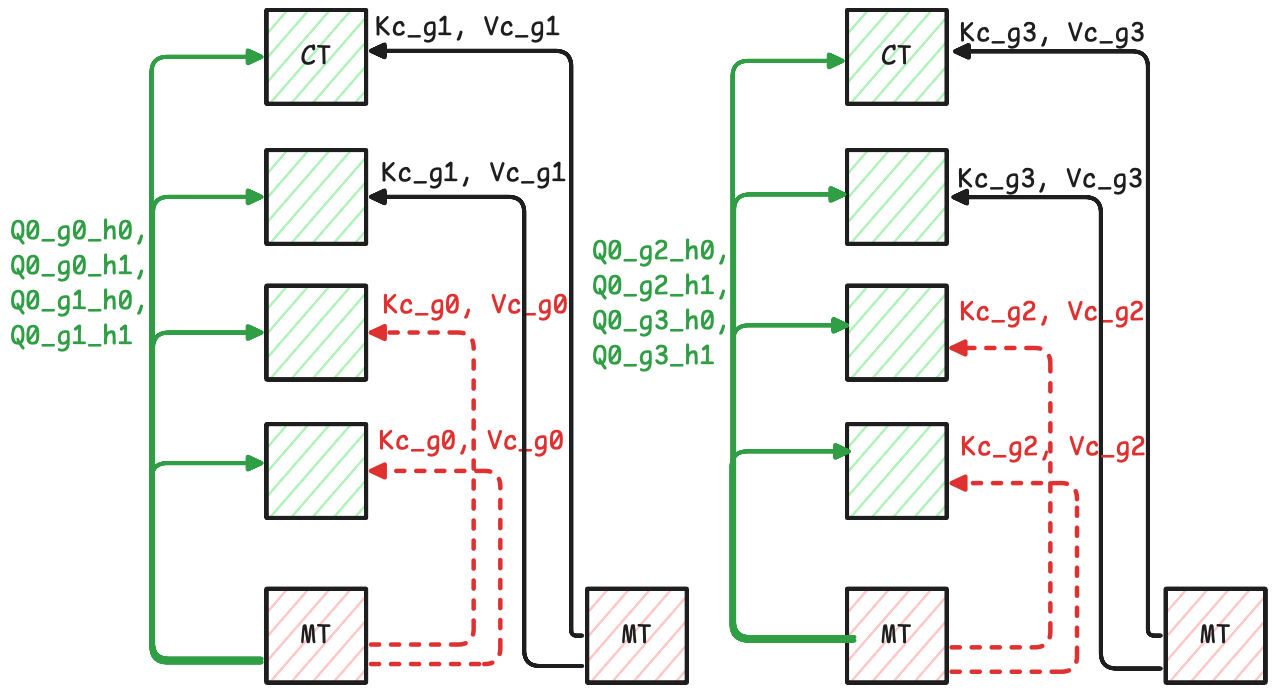} % Replace 'example-image' with your image file name
\caption{
Illustration of FlowQKV execution on an NPU, showing 2 CT columns for $Q_0$ chunk.
}
\label{fig:FlowQKV_npu}
\end{figure}

As shown in Fig.~\ref{fig:FlowQKV_npu}, FlowQKV maps cleanly to the NPU: Gemma3-4B (text) uses 4 KV groups, each serving 2 query heads. Each KV group is handled by 2 adjacent CTs in a col. Queries run in chunks--2 CT columns per Q-chunk--so 8 columns cover 4 chunks (a \textit{super chunk}; see Fig.~\ref{fig:chunked_attention}).

$K$, $V$ chunks (\( \mathbf{Q}_c \)) are loaded into the CTs via two parallel DMA channels, while $Q$ chunks is broadcast to all CTs. Each CT then selects the relevant portion of the $Q$ heads it needs for computation. This enables simultaneous data transfer. For each set of query chunks, a \textit{KV chunk sweep} is performed, where the queries are evaluated against all available KV chunks to compute the attention output. After this sweep completes, the next group of query chunks can be processed. In causal attention (this example), the sweep must span progressively more chunks (or super-chunks) as the context grows. FlowQKV enables attention computation to scale to long contexts with high efficiency, leveraging Ryzen AI's tiled compute architecture and bandwidth hierarchy.

% Similar to the projection layer, we evaluate the performance of FlowQKV using the metric of \textit{read memory bandwidth utilization}, denoted as \( U^{\text{rd}}_{\text{mem}} \). This metric quantifies how effectively the NPU utilizes memory bandwidth during the attention computation phase.

% The read memory bandwidth utilization for FlowQKV is defined as:

% \begin{equation}
% U_{\text{mem}}^{rd} = \frac{L_c \cdot d \cdot 2B \cdot 2}{t_d} \times \frac{\text{CT}_{\text{count}} \cdot H}{2}
% \label{eq:flowqkv_umem}
% \end{equation}

% \noindent where:
% \begin{itemize}
%   \item \( L_c \): chunk size (number of tokens per chunk),
%   \item \( d \): head dimension,
%   \item \( B \): bytes per element (e.g., 2 bytes for bf16),
%   \item \( t_d \): compute time for one chunk (tile latency),
%   \item \( \text{CT}_{\text{count}} \): number of CTs used in parallel
% \end{itemize}

We can use Eq.~\ref{eq:flowqkv_umem_measured} to measure the $U_{\text{mem}}^{rd}$ in runtime. \(T_d\) denotes the total measured runtime over prefill sequence length ($L$). This captures the quadratic \(O(L^2)\) behavior related to attention's inherent complexity.

\begin{equation}
U_{\text{mem}}^{rd} = \frac{2B \cdot \bigl(1 + \tfrac{L}{L_c}\bigr) \cdot L \cdot \mathrm{CT}_{\mathrm{count}} \cdot H}{T_d}
\label{eq:flowqkv_umem_measured}
\end{equation}

A higher \( U_{\text{mem}}^{rd} \) indicates better utilization of the available memory bandwidth. Optimizing FlowQKV to maximize \( U_{\text{mem}}^{rd} \) ensures the design is memory-bound (peak efficiency). This metric is used assess the impact of various tile configurations and scheduling strategies on hardware performance.

FlowQKV has two variants: FlowQKV-SWA (for local layers), which restricts the chunk sweep to a sliding window \((L_c = 1024)\), and FlowQKV-NCA (non-causal for ViT), which sweeps the full sequence. FlowQKV, FlowQKV-SWA, and FlowQKV-NCA share the same hardware configuration; only the instruction schedules differ.

\subsection{Decoding}

The core computations in LLM decoding—projection and attention—are both memory-bound, meaning their performance is limited by \( U_{\text{mem}}^{rd} \). Maximum throughput is achieved when this bandwidth is saturated. 

% We introduces two techniques to address this: FusedDQP (Fused dequantization and projection) for projection operations and FlowKV for attention.

\subsubsection{FusedDQP (Fused dequantization and projection)}

We describe the MVM operation as
\begin{equation}
\mathbf{Y} = \mathbf{W} \times \mathbf{A},
\label{eq:mvm}
\end{equation}
where \( \mathbf{W} \in \mathbb{R}^{M \times K} \) is the projection matrix and \( \mathbf{A} \in \mathbb{R}^{K \times 1} \) is the input activation vector.

\begin{figure}[h!]
    \centering
    \includegraphics[width=0.36\textwidth]{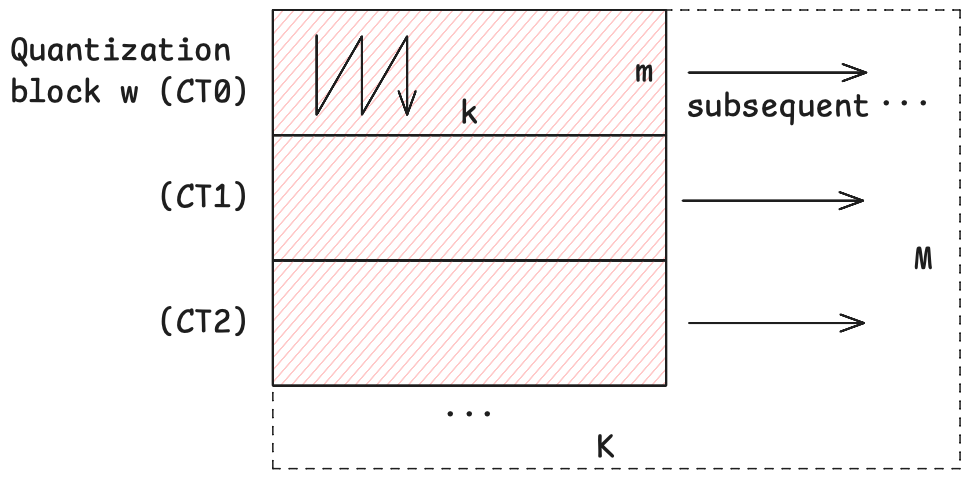} % Replace 'example-image' with your image file name
    \caption{Tiling strategy for FusedDQP: the weight matrix \( \mathbf{W} \) is partitioned into Q4NX-aligned blocks.}
    \label{fig:FusedDQP}
    % \Description{Tiling for FusedDQP}
\end{figure}

We tile \( \mathbf{W} \) into smaller blocks of size \( m \times k = 32 \times 256 \) (Fig.~\ref{fig:FusedDQP}), denoted as \( \mathbf{w} \), which matches the quantization block size. Similarly, \( \mathbf{A} \) is divided into small vectors \( \mathbf{a} \in \mathbb{R}^{k \times 1} \), with \( k = 256 \) chosen to align with Q4NX quantization. This naturally partitions the output \( \mathbf{Y} \) into tiles \( \mathbf{y} \in \mathbb{R}^{m \times 1} \). In the single compute tile (CT) scenario, the first vector \( \mathbf{a} \) is loaded into the CT, and accumulation proceeds via:
\begin{equation}
\mathbf{y}_{\text{acc}} \mathrel{+}= \text{dequant}(\mathbf{w}) \times \mathbf{a}
\label{eq:fusedmv}
\end{equation}

Importantly, the dequantization in Eq.~\ref{eq:fusedmv} does not operate on the entire \( \mathbf{w} \) block at once. Instead, \( \mathbf{w} \) is further partitioned into fine-grained sub-blocks of size \( 16 \times 8 \), which are streamed into vector processor. For each sub-block, dequantization and MVM are executed in a fused kernel, involving only one load from and one store to L1 memory. This fusion maximizes register reuse and minimizes memory access overhead.

Once a \( \mathbf{w} \) is fully used, subsequent \( \mathbf{a} \) vectors and their corresponding \( \mathbf{w} \) blocks (to the right of the previous \( \mathbf{w} \)) are fetched and processed until the full \( \mathbf{y}_{\text{acc}} \) is computed (Fig.~\ref{fig:FusedDQP}). When multiple CTs are used, the vector \( \mathbf{a} \) is broadcast to all CTs. Each CT independently fetches its associated quantized \( \mathbf{w} \) block from DRAM and performs MVM in parallel.

\subsubsection{FlowKV}
FlowKV is an hardware-aware attention scheme that optimizes KV-cache access to maximize DRAM bandwidth.

During LLM decoding, the full KV cache grows with sequence length, creating pressure on memory bandwidth and on-chip storage. To mitigate this, we adopt a chunk-wise processing approach: the sequence is divided into chunks of size \( L_c \), and the NPU processes one chunk at a time. Assuming GQA, each KV head is shared by \( H/G \) query heads. As an example, consider single-head causal full attention. The equations are similar to FlowQKV in prefill (Sec.~\ref{sec:FlowQKV}), except that the Q chunk size is 1 during decoding.

\subsubsection{Efficient Mapping to the NPU}

\begin{figure}[!ht]
    \centering
    \includegraphics[width=0.30\textwidth]{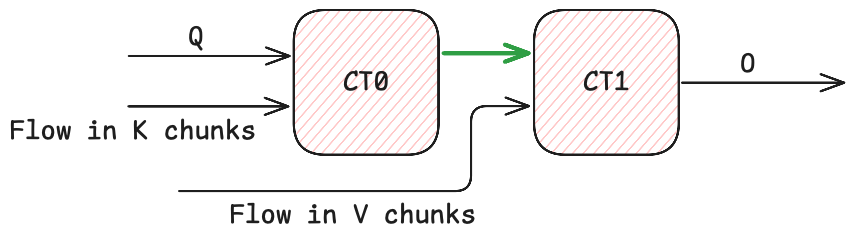} % Replace 'example-image' with your image file name
    \caption{FlowKV pipeline (2-CT scenario): CT0 streams in key chunks to compute attention scores and passes intermediate results ($F$, $C$, $L$) to CT1. CT1 streams in value chunks and performs softmax-weighted accumulation to produce the final output. This pipeline enables concurrent key and value processing, maximizing NPU bandwidth utilization.}
    % \Description{CT0 streams in key chunks to compute attention scores and passes intermediate results ($F$, $C$, $L$) to CT1. CT1 streams in value chunks and performs softmax-weighted accumulation to produce the final output. This pipeline enables concurrent key and value processing, maximizing NPU bandwidth utilization.}
    \label{fig:FlowKV}
\end{figure}

As shown in Fig.~\ref{fig:FlowKV}, two CTs are used to implement the FlowKV attention pipeline. CT0 handles steps (1)–(5) of the FlowKV procedure (Sec.~\ref{sec:FlowQKV}). CT1 performs steps (6)–(7). Intermediate results ($F$, $C$, and $l$) are passed directly from CT0 to CT1 via AIE on-chip memory (green arrow in Fig.~\ref{fig:FlowKV}). Both key and value chunks are streamed into the CTs using double buffering.

\begin{figure}[!ht]
    \centering
    \includegraphics[width=0.48\textwidth]{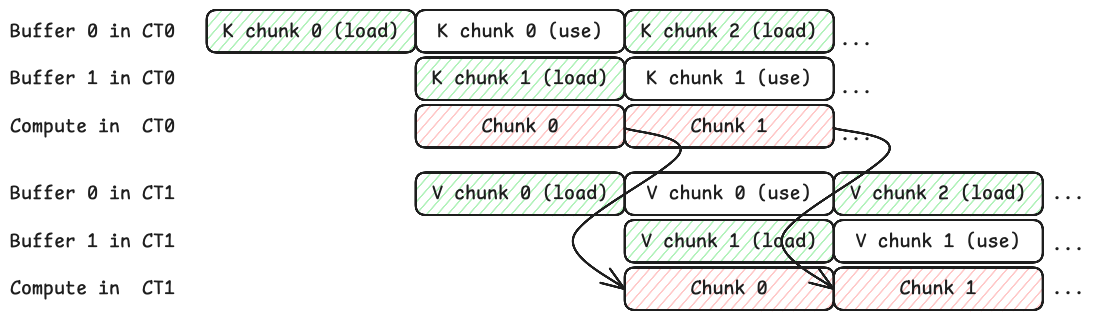}
    \caption{Timing diagram of the FlowKV pipeline (2-CT configuration) under ideal conditions, where data transfer is fully overlapped with kernel execution in both CTs.}
    \label{fig:FlowKV_timing}
    % \Description{Timing_FlowKV}
\end{figure}

Ideally, the computation latency for each chunk overlaps with its transfer time, effectively hiding kernel execution overhead, illustrated in Fig.~\ref{fig:FlowKV_timing}. While Fig.~\ref{fig:FlowKV} illustrates a 2-CT example, the design is flexible: the chunk size \( L_c \), number of CTs, or number of KV heads can be adjusted to data transfer time and compute intensity towards ideal case to achieve the best performance.

Two CT columns (8 CTs) are assigned to the FlowKV kernel. In this configuration, every two CTs operate as a pair (as illustrated in Fig.~\ref{fig:FlowKV}). Thus, a total of 4 pairs are formed. One KV head (a group) mapped to each CT pair. This mapping balances the chunk data transfer time with kernel execution latency, enabling efficient pipelined execution across the CTs and maximizing \( U_{\text{mem}}^{rd} \). For projection operations, four CT columns (16 CTs) to the FusedDQP kernel. One CT is dedicated to each of the nonlinear operations: RoPE, RMS normalization with residual connection, and GeLU activation with MUL (Fig.~\ref{fig:Gemma3_4B}). These functions operate on vector inputs, which are temporarily stored in MTs between compute stages. All nonlinear kernels are fully pipelined, and their execution incurs negligible latency compared to projection and attention operations during LLM decoding.

FlowKV has 1 variant: FlowKV-SWA (for local layers), which limits the chunk sweep to a sliding window ($L_c = 1024$). FlowKV and FlowKV-SWA share the same hardware configuration; only the instruction differs.

\section{Experimental Results}

\subsection{Implementation}

We implemented Gemma3 models (1B and 4B variants, where 4B has the vision tower) on the AMD Ryzen AI 7 350 NPU, integrated into the ASRock 4X4 BOX-AI350 mini-PC with SO-DIMM DDR5 5600 MHz \cite{amd2025ryzen350}. This compact system features a ``Kraken Point'' APU with an XDNA 2-based NPU.

The full implementation and low-level kernel optimization are carried out using AMD’s AIE-MLIR infrastructure along with the IRON Python interface \cite{mlir_aie, hunhoff2025efficiency}. Link: (\github{https://github.com/FastFlowLM/FastFlowLM})

\subsection{Evaluation and Benchmarking Results}

We benchmarked our proposed method's performance on both prefill and decoding. For prefill speed (TTFT, Time To First Token), we evaluated 1B and 4B model variants from 1K to 32K tokens, shown in Tables \ref{tab:ttft_gemma3_1b} and \ref{tab:ttft_gemma3_4b}. For decoding speed (TPS, Token Per Second), we tested models across their full supported range, from 1K up to 32K tokens for the 1B variant and up to 128K (131,072) tokens for the 4B variant, shown in Table \ref{tab:decode_gemma3_1b} and \ref{tab:decode_gemma3_4b}.

Additionally, we conducted benchmarks using the iGPU (LM Studio \cite{lmstudio2025}) and the CPU (Ollama \cite{ollama}), all on the same processor. 

% Prefill (TTFT, ms) up to 32k, transposed by hardware

\begin{table}[h]
\centering
\small
\caption{Gemma3-1B — Prefill TTFT (sec)}
\begin{tabular}{lrrrrrr}
\toprule
\textbf{Hardware} & \textbf{1k} & \textbf{2k} & \textbf{4k} & \textbf{8k} & \textbf{16k} & \textbf{32k} \\
\midrule
NPU  & 0.95 & 1.47 & 2.46  & 4.42  & 8.45 & 17.19 \\
iGPU & 0.51 & 1.14 & 2.61  & 6.64 & 21.3 & 95.1 \\
CPU  & 4.06 & 8.06 & 17.2 & 36.2 & 75.5 & 165 \\
\bottomrule
\end{tabular}
\label{tab:ttft_gemma3_1b}
\end{table}

\begin{table}[h]
\centering
\small
\caption{Gemma3-4B — Prefill TTFT (sec)}
\begin{tabular}{lrrrrrr}
\toprule
\textbf{Hardware} & \textbf{1k} & \textbf{2k} & \textbf{4k} & \textbf{8k} & \textbf{16k} & \textbf{32k} \\
\midrule
NPU  & 1.81 & 2.81 & 4.79 & 8.37 & 16.17 & 33.5 \\
iGPU & 2.05 & 4.26 & 9.54 & 23.8 & 71.5 & 265  \\
CPU  & 20.3 & 42.4 & 85.4 & 176  & 766  & 832  \\
\bottomrule
\end{tabular}
\label{tab:ttft_gemma3_4b}
\end{table}

% Decoding (TPS) up to 32k — rounded to ~3 significant digits

\begin{table}[h]
\centering
\small
\caption{Gemma3-1B — Decoding Throughput (TPS)}
\begin{tabular}{lrrrrrr}
\toprule
\textbf{Hardware} & \textbf{1k} & \textbf{2k} & \textbf{4k} & \textbf{8k} & \textbf{16k} & \textbf{32k} \\
\midrule
NPU  & 41.1 & 40.5 & 39.5 & 37.3 & 33.6 & 27.9 \\
iGPU & 38.0 & 53.9 & 42.3 & 33.5 & 25.0 & 13.6 \\
CPU  & 41.9 & 40.8 & 41.7 & 40.2 & 38.1 & 33.8 \\
\bottomrule
\end{tabular}
\label{tab:decode_gemma3_1b}
\end{table}

% Compact layout: tighter column spacing & row height, 3-digit style
\begingroup
\setlength{\tabcolsep}{3.5pt}
\renewcommand{\arraystretch}{0.9}

\begin{table}[h]
\centering
\small
\caption{Gemma3-4B — Decoding Throughput (TPS)}
\begin{tabular}{lrrrrrrrr}
\toprule
\textbf{HW} & \textbf{1k} & \textbf{2k} & \textbf{4k} & \textbf{8k} & \textbf{16k} & \textbf{32k} & \textbf{64k} & \textbf{128k} \\
\midrule
NPU  & 18.2 & 18 & 17.8 & 17.3 & 16.3 & 14.8 & 13.2 & 11.2 \\
iGPU & 18.6 & 17.4 & 15.3 & 12.6 & 9.2 & 5.9 & 3.3 & 1.9 \\
CPU  & 14.6 & 13.5 & 13.9 & 13.0 & 11.4 & 10.8 & 7.5 & 4.1 \\
\bottomrule
\end{tabular}
\label{tab:decode_gemma3_4b}
\end{table}
\endgroup

For vision-tower inference, NPU TTFT is 2.6s vs 7.45s (iGPU) and 38.55s (CPU), yielding \(2.9\times\) and \(14.8\times\) speedups, respectively.

We used HWiNFO to monitor all test cases, capturing average chip temperature and the power consumption of the CPU, iGPU, and NPU; note that board-level draw (e.g., memory controllers and other motherboard components) is not included in these power figures. Under iGPU workloads (1B and 4B, prefill and decoding), the chip reached ~98 °C within 5 sec; under CPU workloads (1B and 4B, prefill and decoding), it reached ~100.5 °C within 5 sec; with the NPU, the chip temperature remained consistently below 50 °C.

% Cross-Hardware Power Consumption (W) with Totals — compact, multirow
% Requires: \usepackage{booktabs,multirow}
\begingroup
\setlength{\tabcolsep}{6pt}
\renewcommand{\arraystretch}{0.9}

\begin{table}[h]
\centering
\small
\caption{Cross-Hardware Power Consumption Average (W)}
\begin{tabular}{l l r r r r}
\toprule
\textbf{Category} & \textbf{Model} & \textbf{CPU} & \textbf{GPU} & \textbf{NPU} & \textbf{Total} \\
\midrule
\multirow{2}{*}{NPU Decoding} & 1B & 2.8 & 0   & 1.8 & 4.6 \\
                              & 4B & 2.9 & 0   & 1.6 & 4.5 \\
\addlinespace
\multirow{2}{*}{NPU Prefill}  & 1B & 0.0 & 3.1 & 1.2 & 4.3 \\
                              & 4B & 0.0 & 3.4 & 1.1 & 4.5 \\
\addlinespace
\multirow{2}{*}{iGPU Decoding} & 1B & 31  & 22  & 0   & 53  \\
                              & 4B & 31  & 23  & 0   & 54  \\
\addlinespace
\multirow{2}{*}{iGPU Prefill}  & 1B & 33  & 24  & 0   & 57  \\
                              & 4B & 33  & 25  & 0   & 58  \\
\addlinespace
\multirow{2}{*}{CPU Decoding} & 1B & 29  & 0   & 0   & 29  \\
                              & 4B & 26  & 0   & 0   & 26  \\
\addlinespace
\multirow{2}{*}{CPU Prefill}  & 1B & 24  & 0   & 0   & 24  \\
                              & 4B & 30  & 0   & 0   & 30  \\
\bottomrule
\end{tabular}
\label{tab:power_totals}
\end{table}

\endgroup

\begin{figure}[htbp]
  \centering
  
  \begin{subfigure}
    \centering
    \includegraphics[width=0.8\linewidth]{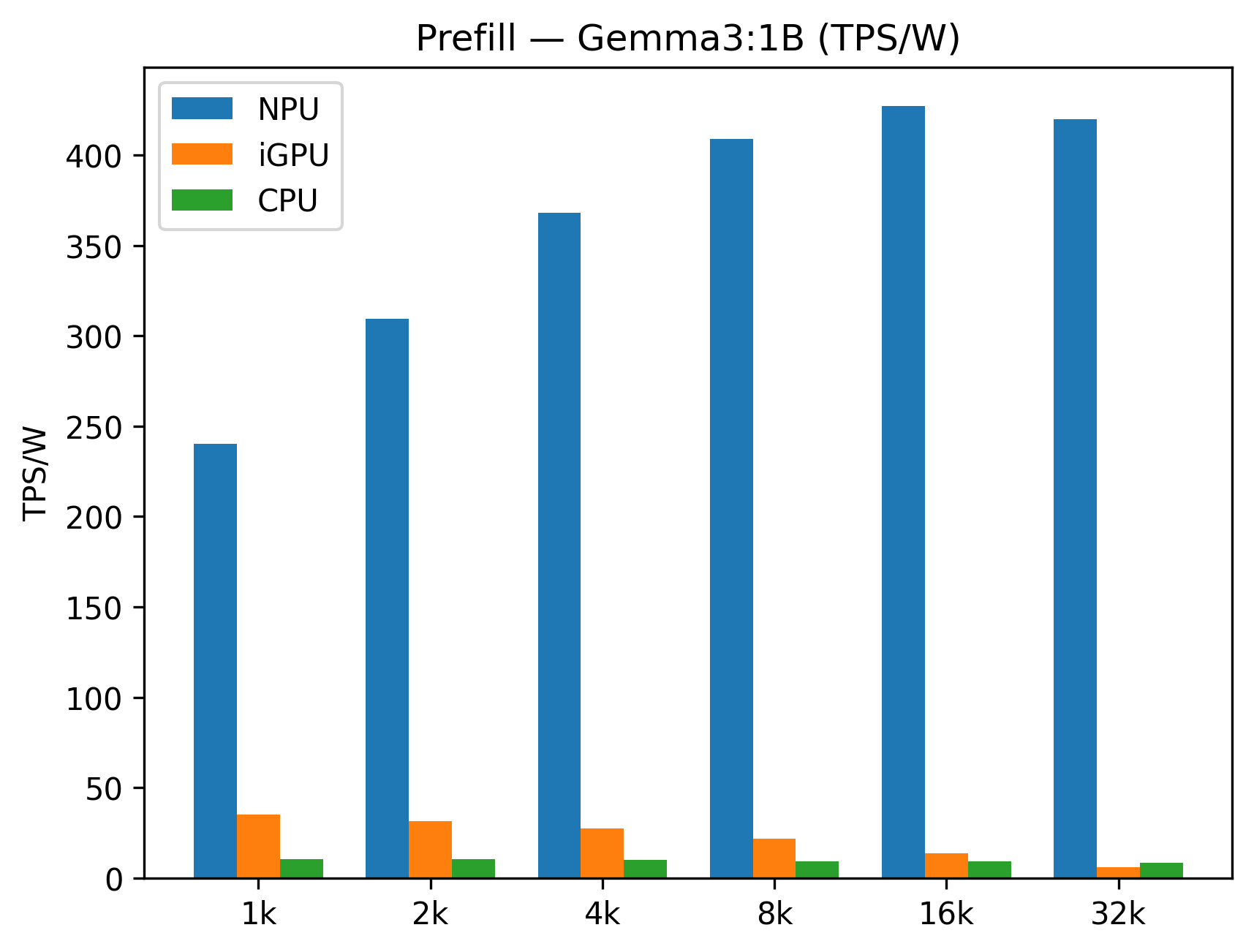}
    \label{fig:prefill_1b}
  \end{subfigure}
  
  \begin{subfigure}
    \centering
    \includegraphics[width=0.8\linewidth]{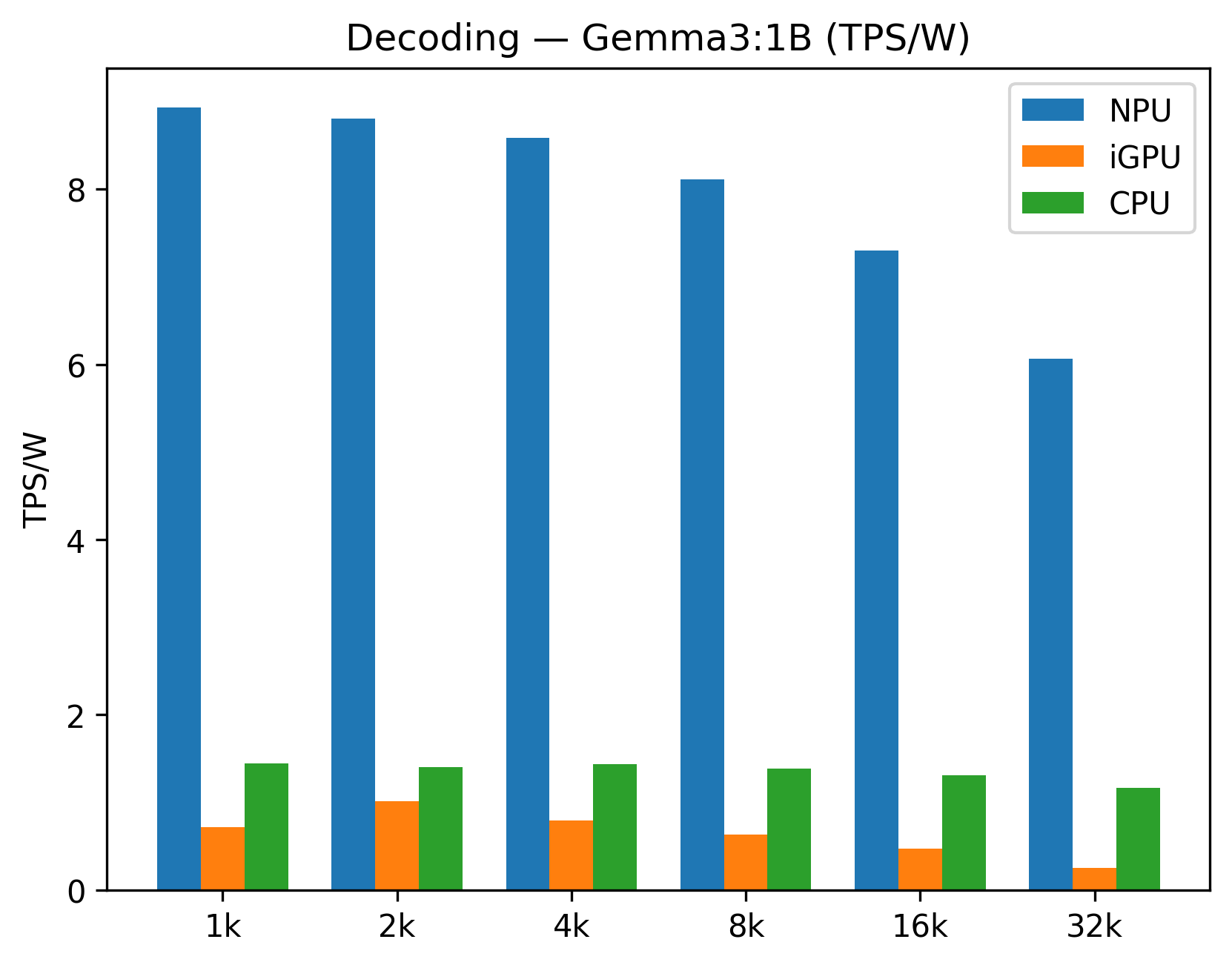}
    \label{fig:decode_1b}
  \end{subfigure}

  \begin{subfigure}
    \centering
    \includegraphics[width=0.8\linewidth]{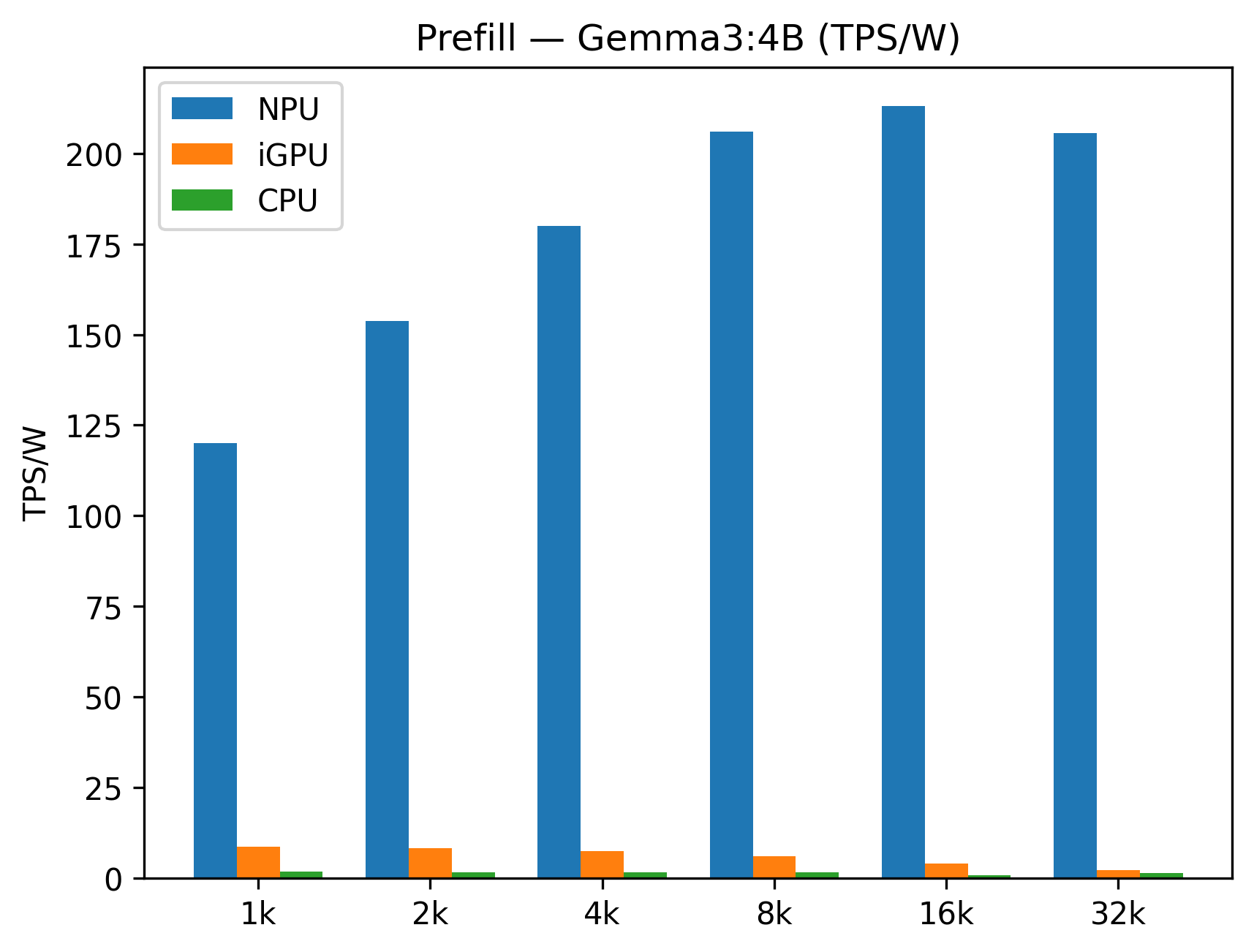}
    \label{fig:prefill_4b}
  \end{subfigure}
 
  \begin{subfigure}
    \centering
    \includegraphics[width=0.8\linewidth]{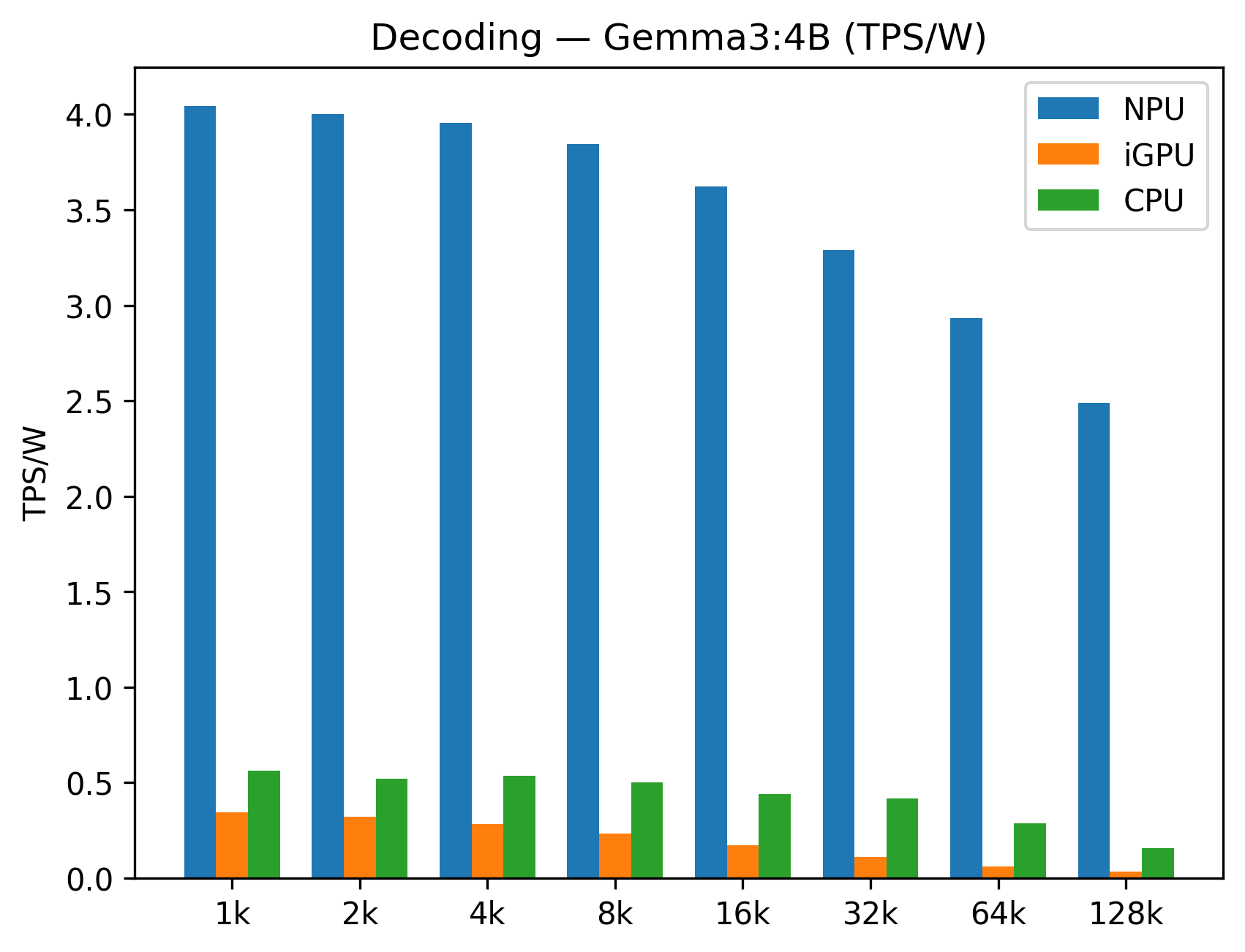}
    \label{fig:decode_4b}
  \end{subfigure}

  \caption{Energy efficiency (TPS/W) across sequence lengths for prefill and decoding on Gemma3 1B and 4B. Colors are consistent across plots: NPU (green), iGPU (orange), CPU (red).}
  \label{fig:tpsw_gemma3_all}
\end{figure}

TPS/watt was calculated to show the power efficiency of all cases at different sequence lengths, shown in Fig. \ref{fig:tpsw_gemma3_all}.

To summarize, for prefill, the 1B variant shows NPU power-efficiency gains of \(6.9\times\)–\ \(69.7\times\) over iGPU and \(22.9\times\)–\ \(50.7\times\) over CPU across 1k–32k tokens; the 4B variant delivers \(13.9\times\)–\ \(96.7\times\) over iGPU and \(70.6\times\)–\ \(157.7\times\) over CPU across 1k–32k. For decoding, the 1B variant achieves \(12.4\times\)–\ \(24.2\times\) over iGPU and \(6.2\times\)–\ \(5.2\times\) over CPU across 1k–32k, while the 4B variant yields \(11.9\times\)–\ \(71.1\times\) over iGPU and \(7.2\times\)–\ \(15.8\times\) over CPU across 1k–128k.

\section{Discussion}

% The proposed method runs the exact same LLM model without any algorithmic modifications, ensuring that accuracy remains unchanged. Both FusedDQP and FlowKV maintain an effective read memory bandwidth utilization of 32 GB/s (constant, not average), assuming running the NPU at 1 GHz (Bandwidth utilization is linearly proportional to the clock rate). It is worth noting that the bandwidth allocation to the NPU, depends on the hardware vendor’s unified memory configuration and the specific DRAM type used. When the available memory read bandwidth is less than 32 GB/s, the system becomes memory-bound. Conversely, when sufficient bandwidth is available, performance becomes compute-bound. The proposed design can work on all AMD Ryzen AI XDNA2 NPUs. Notably, our solution is linearly scalable with the number of CTs. For example, increasing total CT columns from 8 to 12 (i.e., adding 16 CTs) is expected to raise the effective read memory bandwidth utilization of both FusedDQP and FlowKV to 48 GB/s.

Integrated NPUs are designed for low-power, privacy-preserving, always-on edge scenarios such as laptops and handheld devices, rather than multi-tenant datacenter workloads or very large models; they are not intended to compete directly with discrete or datacenter GPUs in that regime.

The proposed method executes unmodified LLMs, preserving full model accuracy without any algorithmic changes. For MM, FusedDQP (and its variants), and FlowKV/FlowQKV (and their variants), the theoretical sustained read bandwidth requirement \(U_{\text{mem}}^{\mathrm{rd}}\) exceeds \(60\,\mathrm{GB/s}\) at \(1.8\,\mathrm{GHz}\) and scales linearly with clock frequency. A system is memory-bound when the vendor-allocated NPU read bandwidth is lower than \(U_{\text{mem}}^{\mathrm{rd}}\); otherwise it is compute-bound.

On the test chip, the NPU's read memory bandwidth is capped below \(40\,\mathrm{GB/s}\), so these kernels operate in the memory-bound regime. The CPU and iGPU get far more memory bandwidth, which explains their decoding speed advantage at short contexts. Under sufficient bandwidth, MM, FusedDQP,  FlowKV/FlowQKV scale linearly with the number of compute tiles (CTs). As future NPUs provide higher memory bandwidth and more CTs, they are expected to substantially outperform CPU- and iGPU-based solutions in both speed and efficiency.

We are also developing and maturing toolchains and libraries to accelerate LLM deployment on the NPU chips; these will be released in the near future.

Although demonstrated on AMD Ryzen AI NPUs, we believe the techniques generalize to dataflow-style accelerators with tiled compute, local SRAM, and DMA or broadcast, and can be adapted to platforms such as Cerebras, and AMD AIE/ACAP, etc.

\section{Conclusion}
We demonstrate the first end-to-end mapping of Gemma3 text--vision models onto a tiled edge NPU (AMD Ryzen AI NPU), showing that hardware-aware scheduling and bandwidth-centric kernels enable efficient, accurate, low-power inference without algorithmic changes. FlowQKV and FlowKV sustain high DRAM utilization for attention, FusedDQP removes projection overheads, and the compact Q4NX format reduces memory footprint while preserving fidelity. On Ryzen AI, these techniques yield multi-\(\times\) speedups and up to \(150\times\)+ power-efficiency gains over CPU/iGPU baselines. The principles generalize to future NPUs and other dataflow accelerators, enabling scalable, real-time LLM/VLM inference at the edge.

\bibliography{ref}
\bibliographystyle{mlsys2025}

%%%%%%%%%%%%%%%%%%%%%%%%%%%%%%%%%%%%%%%%%%%%%%%%%%%%%%%%%%%%%%%%%%%%%%%%%%%%%%%
%%%%%%%%%%%%%%%%%%%%%%%%%%%%%%%%%%%%%%%%%%%%%%%%%%%%%%%%%%%%%%%%%%%%%%%%%%%%%%%
% SUPPLEMENTAL CONTENT AS APPENDIX AFTER REFERENCES
%%%%%%%%%%%%%%%%%%%%%%%%%%%%%%%%%%%%%%%%%%%%%%%%%%%%%%%%%%%%%%%%%%%%%%%%%%%%%%%
%%%%%%%%%%%%%%%%%%%%%%%%%%%%%%%%%%%%%%%%%%%%%%%%%%%%%%%%%%%%%%%%%%%%%%%%%%%%%%%
% \appendix
% \section{Please add supplemental material as appendix here}
% %
% Put anything that you might normally include after the references as an appendix here, {\it not in a separate supplementary file}. Upload your final camera-ready as a single pdf, including all appendices.

%%%%%%%%%%%%%%%%%%%%%%%%%%%%%%%%%%%%%%%%%%%%%%%%%%%%%%%%%%%%%%%%%%%%%%%%%%%%%%%
%%%%%%%%%%%%%%%%%%%%%%%%%%%%%%%%%%%%%%%%%%%%%%%%%%%%%%%%%%%%%%%%%%%%%%%%%%%%%%%

\end{document}